\documentclass[aps,prl,twocolumn,superscriptaddress,showpacs]{revtex4}
\usepackage{amssymb}
\usepackage{amsmath}
\usepackage{graphicx}
\usepackage{makeidx}
\usepackage{amsfonts}
\usepackage{bm}
\usepackage{color}

\input{tcilatex}

\begin{document}

\title{Prediction of termomagnetic and termoelectric properties for novel
materials and systems}
\author{A.A. Varlamov}
\affiliation{CNR-SPIN, Viale del Politecnico 1, I-00133 Rome, Italy}
\author{Alexey V. Kavokin}
\affiliation{Physics and Astronomy School, University of Southampton, Highfield,
Southampton, SO171BJ, UK and Laboratoire Charles Coulomb, CNRS-Universite de
Montpellier II, Pl. Eugene de Bataillon, 34095 Montpellier Cedex, France}
\date{\today }

\begin{abstract}
We express the link between conductivity and coefficients of Seebeck,
Nernst-Ettingshausen, Peltier, and Thompson and Reghi-Leduc via the
temperature derivative of the chemical potential of a system. These general
expressions are applied to three-, two- and one-dimensional systems of
charge carriers having a parabolic or Dirac spectrum. The method allows for
predicting thermoelectric and thermomagnetic properties of novel materials
and systems.
\end{abstract}

\pacs{ 72.15.Jf, 72.20.Pa }
\maketitle

The theory of thermoelectric (TE) and thermomagnetic (TM) phenomena in
metals has been built in 1930s-1950s \cite{Mott,1948Sondheimer}.
Essentially, it is based on the kinetic approach, where more or less
complicated transport equations are formulated and solved for different
systems in order to obtain the transport coefficients characterizing the TE
and TM effects. In the recent decades, invention of a wide range of new
materials with exotic spectra where different types of interactions can
interplay (graphene and carbon nano-tubes being two examples) gave a boost
to the studies of the most important TE and TM constants, such as Seebeck,
Thomson, Nernst-Ettingshausen and Reghi-Leduc coefficients, thermal
conductivity and Peltier tensors. Yet, the notion of a heat flow, required
to find these coefficients, becomes hardly definable in the case of systems
of interacting particles, which is why one can hardly rely on kinetic
approaches, in general. Such a problem does not appear if one deals with the
conductivity tensor which can be always calculated using either transport
equations or diagrammatic approaches. Some relations between the TE and TM
constants and conductivity tensor are well known for non-interacting systems
with simple spectra (Wiedemann-Franz law\&\ Mott formula), but these
relations have not been generalised to the case of interacting systems with
exotic spectra so far.

This work is aimed at formulating a unified approach to description of TE
and TM phenomena virtually in any electronic system based on establishing
the universal links between main TE and TM coefficients and the conductivity
tensor\textit{. }We show that it is sufficient to know the temperature
dependence of a chemical potential of a system to obtain Seebeck, Thomson
and Peltier coefficients. The Nernst-Ettingshausen, Reghi-Leduc and thermal
conductivity coefficients can be expressed through the conductivity tensor
and thermal derivatives of the chemical potential and magnetization of the
system. These relations allow for obtaining the thermoelectric and
thermomagnetic properties of novel 1D and 2D systems of normal charge
carriers and Dirac fermions, electron systems with topologically nontrivial
spectra etc.

We shall operate with the chemical and electro-chemical potentials of charge
carriers in a wide range of systems and materials. What a wonder, these
quantities have a non-unique definition in literature! Electro-chemists and
soft-matter physicists usually assume that an electro-chemical potential of
a system is a constant in a stationary conditions, while a chemical
potential is a local characteristic which may change from point to point of
a system (see, for example, \cite{Bard}). In the solid state physics,
frequently, the opposite rule is postulated: a chemical potential is a
characteristic of the whole system, and it is a constant in the stationary
case, while the electrochemical potential may vary from point to point (see,
for example, \cite{Ashkroft}).

Here we shall adopt the former approach, following the textbooks of Madelung 
\cite{Madelung} and Abrikosov \cite{Abrikosov}, who applied the concept of
local chemical potential to solid state systems. In this approach, the
system subjected to a temperature gradient is assumed to be in thermal
equilibrium locally, so that in each small volume of the sample one can
introduce the thermodynamic potential $\Omega \left[ T\left( \mathbf{r}%
\right) \right] $, $\mathbf{r}$ being a coordinate, the number of particles $%
\ N\left[ T\left( \mathbf{r}\right) \right] $ and the chemical potential 
\begin{equation}
\mu \left[ T\left( \mathbf{r}\right) \right] =-\frac{\partial \Omega }{%
\partial N},  \label{1}
\end{equation}
Defined in this way, the chemical potential may vary in real space, if the
temperature of the system varies.

The electrochemical potential is defined as 
\begin{equation}
\overline{\mu }=\mu +e\varphi ,  \label{2}
\end{equation}
with $\varphi $ being the electrostatic potential. This quantity remains
constant for a whole system at stationary conditions. Physically it means
that if no electric current flows through the system, its electro-chemical
potential is constant, while its chemical potential can vary.

The temperature dependencies of chemical potentials for normal carriers
(having a parabolic dispersion) and Dirac fermions (having a linear
dispersion) for the systems of different dimensionalities in the Boltzmann
limit and in the limit of a degenerate Fermi gas are summarized in Table 1.
We show below that this information is sufficient for predication TE and TM
coefficients in a very wide variety of systems. 
\begin{figure}[t]
\includegraphics[width=.9\columnwidth]{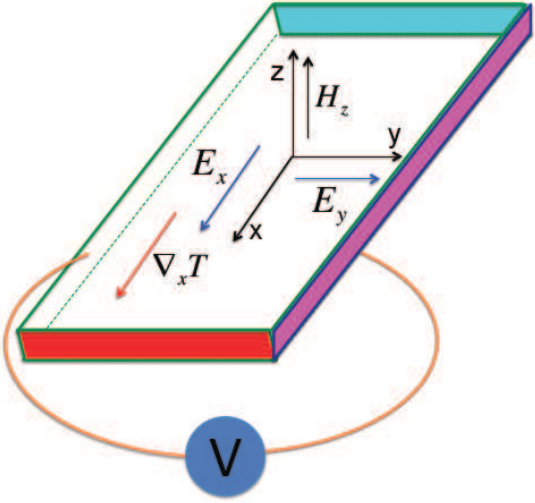}
\caption{Schematic: geometry of the experiments considered in this paper. }
\label{geometry}
\end{figure}
Let us consider a conductor looped via a voltmeter in y-direction, placed in
the magnetic field $\mathbf{H}$ oriented along z-axis, and\ subjected to the
temperature gradient $\nabla _{x}T$ \ applied along x-axis (see Fig. \ref%
{geometry}). In a full generality, one can express the electric current
density $\mathbf{j}$ components as:%
\begin{equation}
\left( 
\begin{array}{c}
j_{x} \\ 
j_{y}%
\end{array}%
\right) =\widehat{\sigma }\left( 
\begin{array}{c}
E_{x} \\ 
E_{y}%
\end{array}%
\right) +\widehat{\beta }\left( 
\begin{array}{c}
\nabla _{x}T \\ 
0%
\end{array}%
\right) ,  \label{j}
\end{equation}%
where $\widehat{\sigma }$ and $\widehat{\beta }$ are conductivity and
thermoelectric tensors, respectively. Further we restrict our consideration
to the limit of magnetic fields weak by a parameter $\omega _{c}\tau \ll 1$,
where $\omega _{c}$ is the cyclotron frequency, and $\tau $ is the elastic
scattering time.\ 

\begin{widetext}
$%
\begin{tabular}{|l|l|l|l|}
\hline
Dimensionality & $d=3$ & $d=2$ & $d=1$ \\ \hline
\begin{tabular}{l}
Fermi energy $\mu _{P}^{\left( d\right) }\left( 0\right) $ \\
(parabolic spectrum)%
\end{tabular}%
$\ $ & $\mu _{P}^{\left( 3\right) }\left( 0\right) =\left( 3\pi ^{2}\right)
^{2/3}\frac{\hbar ^{2}}{2m}\left[ n_{e}^{\left( 3\right) }\right] ^{2/3}$ & $%
\mu _{P}^{\left( 2\right) }\left( 0\right) =\frac{\pi \hbar ^{2}}{m}%
n_{e}^{\left( 2\right) }$ & $\mu _{P}^{\left( 1\right) }\left( 0\right) =%
\frac{2\pi ^{2}\hbar ^{2}}{m}\left[ n_{e}^{\left( 1\right) }\right] ^{2}$ \\
\hline
\begin{tabular}{l}
Chemical potential \\
for degenerated FG \\
(parabolic spectrum) \\
$T\ll \mu _{P}^{\left( D\right) }\left( 0\right) $%
\end{tabular}
& $\mu _{P}^{\left( 3\right) }\left( T\right) =\mu _{P}^{\left( 3\right)
}\left( 0\right) -\frac{\pi ^{2}T^{2}}{6\mu _{P}^{\left( 3\right) }\left(
0\right) }$ &
\begin{tabular}{l}
$\mu _{P}^{\left( 2\right) }\left( T\right) =\mu _{P}^{\left( 2\right)
}\left( 0\right) $ \\
$+T\ln \left[ 1-e^{-\frac{\mu _{P}^{\left( 2\right) }\left( 0\right) }{T}}%
\right] $ \\
$\simeq \mu _{P}^{\left( 2\right) }\left( 0\right) -Te^{-\frac{\mu
_{P}^{\left( 2\right) }\left( 0\right) }{T}}$%
\end{tabular}
& $\mu _{P}^{\left( 1\right) }\left( T\right) =\mu _{P}^{\left( 1\right)
}\left( 0\right) -\frac{\pi ^{2}T^{2}}{12\mu _{P}^{\left( 1\right) }\left(
0\right) }$ \\ \hline
\begin{tabular}{l}
Chemical potential \\
for Boltzman FG \\
(parabolic spectrum) \\
$T\gg \mu _{P}^{\left( d\right) }\left( 0\right) $%
\end{tabular}
$\ \ \ \ $ & $\mu _{P}^{\left( 3\right) }\left( T\right) =-\frac{3}{2}T\ln
\frac{T}{\mu _{P}^{\left( 3\right) }\left( 0\right) }$ &
\begin{tabular}{l}
$\mu _{P}^{\left( 2\right) }\left( T\right) =\mu _{P}^{\left( 2\right)
}\left( 0\right) $ \\
$+T\ln \left[ 1-e^{-\frac{\mu _{P}^{\left( 2\right) }\left( 0\right) }{T}}%
\right] $ \\
$\simeq -T\ln \frac{T}{\mu _{P}^{\left( 2\right) }\left( 0\right) }$%
\end{tabular}
& $\mu _{P}^{\left( 1\right) }\left( T\right) =-\frac{T}{2}\ln \frac{\pi
^{4}T}{4\mu _{P}^{\left( 1\right) }\left( 0\right) }$ \\ \hline
\begin{tabular}{l}
Fermi energy $\mu _{D}^{\left( d\right) }\left( 0\right) $ \\
(Dirac spectrum)%
\end{tabular}
& $\mu _{D}^{\left( 3\right) }\left( 0\right) =\pi \hbar c\sqrt[3]{3n_{e}}$
& $\mu _{D}^{\left( 2\right) }\left( 0\right) =\hbar c\sqrt{2\pi n_{e}}$ & $%
\mu _{D}^{\left( 1\right) }\left( 0\right) =\pi \hbar cn_{e}$ \\ \hline
\begin{tabular}{l}
Chemical potential \\
for degenerated FG \\
(Dirac spectrum) \\
$T\ll \mu _{D}^{\left( d\right) }\left( 0\right) $%
\end{tabular}
& $\mu _{D}^{\left( 3\right) }\left( T\right) =\mu _{D}^{\left( 3\right)
}\left( 0\right) -\frac{\pi ^{2}T^{2}}{3\mu _{D}^{\left( 3\right) }\left(
0\right) }$ & $\mu _{D}^{\left( 2\right) }\left( T\right) =\mu _{D}^{\left(
2\right) }\left( 0\right) -\frac{\pi ^{2}T^{2}}{12\mu _{D}^{\left( 2\right)
}\left( 0\right) }$ &
\begin{tabular}{l}
$\mu _{D}^{\left( 1\right) }\left( T\right) =T\ln \left[ e^{\frac{\mu
_{D}^{\left( 1\right) }\left( 0\right) }{T}}-1\right] $ \\
$\simeq \mu _{D}^{\left( 1\right) }\left( 0\right) -Te^{-\frac{\mu
_{D}^{\left( 1\right) }\left( 0\right) }{T}}$%
\end{tabular}
\\ \hline
\begin{tabular}{l}
Chemical potential \\
for Boltzman FG \\
(Dirac spectrum) \\
$T\gg \mu _{D}^{\left( d\right) }\left( 0\right) $%
\end{tabular}
& $\mu _{D}^{\left( 3\right) }\left( T\right) =-3T\ln \frac{T}{\mu
_{D}^{\left( 3\right) }\left( 0\right) }$ & $\mu _{D}^{\left( 2\right)
}\left( T\right) =-2T\ln \frac{T}{\mu _{D}^{\left( 2\right) }\left( 0\right)
}$ &
\begin{tabular}{l}
$\mu _{D}^{\left( 1\right) }\left( T\right) =T\ln \left[ e^{\frac{\mu
_{D}^{\left( 1\right) }\left( 0\right) }{T}}-1\right] $ \\
$\simeq -T\ln \left( \frac{T}{\mu _{D}^{\left( 1\right) }\left( 0\right) }%
\right) $%
\end{tabular}
\\ \hline
\end{tabular}%
$ \bigskip

Table 1. The temperature dependencies of chemical potential in the systems
of different dimensionalities, for carriers having a parabolic and linear
dispersion, in the limits of  Boltzmann and degenerate Fermi gases.
Expressions related to the Boltzmann gas are shown on the grey background.
$P$ and $D$ denote the expressions obtained for parabolic and Dirac dispersion cases, respectively.
\end{widetext}

For the heat flow$\ \mathbf{q}$ components one can write the similar
equation:%
\begin{equation}
\left( 
\begin{array}{c}
q_{x} \\ 
q_{y}%
\end{array}%
\right) =\widehat{\gamma }\left( 
\begin{array}{c}
E_{x} \\ 
E_{y}%
\end{array}%
\right) +\widehat{\zeta }\left( 
\begin{array}{c}
\nabla _{x}T \\ 
0%
\end{array}%
\right) ,
\end{equation}%
where the tensor $\widehat{\gamma }$\ is related to $\widehat{\beta }$\ by
means of Onsager relation: $\widehat{\gamma }(H)=-T\widehat{\beta }(-H)$.
Tensor $\widehat{\zeta }=-\frac{\pi ^{2}T}{3e^{2}}\widehat{\sigma }$\ mainly
determines the value of thermal conductivity $\widehat{\kappa }$ \cite%
{Abrikosov}. In the following we shall find the tensor $\widehat{\beta }$
and express the most important coefficients of the TM and TE transport
through its components and conductivity tensor $\widehat{\sigma }$.

We limit ourselves to consideration of the case\ where the electric circuits
are broken in both x,y-directions, so that $j_{x}=0$ (see Fig. 1): 
\begin{equation}
j_{x}=\sigma _{xx}E_{x}+\sigma _{xy}E_{y}+\beta _{xx}\nabla _{x}T=0
\label{bound1}
\end{equation}%
\begin{equation}
j_{y}=\sigma _{yx}E_{x}+\sigma _{yy}E_{y}+\beta _{yx}\nabla _{x}T=0
\label{bound2}
\end{equation}%
The off-diagonal components of the TE tensor differ from zero only due to
non-zero magnetic field applied. They can be determined from the fourth
Maxwell equation \cite{Obraztsov,VarlamovPRL} and expressed in terms of the
temperature derivative of the magnetization $M_{z}$: 
\begin{equation*}
\beta _{xy}=-\beta _{yx}=c\frac{\partial M_{z}}{\partial T},
\end{equation*}%
where the latter can be expressed as the derivative of the thermodynamic
potential $\Omega $ over magnetic field%
\begin{equation*}
M_{z}=-\frac{\partial \Omega }{\partial H_{z}}.
\end{equation*}%
The electric field along x-direction induced by the temperature gradient can
be determined using the condition of constancy of the electrochemical
potential Eq. (\ref{2}): 
\begin{equation}
\nabla \left[ e\varphi +\mu (T(x),n(x))\right] =0  \label{const}
\end{equation}%
and can be expressed in terms of the full derivative of the chemical
potential: 
\begin{equation}
E_{x}=-\nabla \varphi =\frac{1}{e}\left( \frac{d\mu }{dT}\right) \nabla
_{x}T.  \label{ex}
\end{equation}%
Here we have assumed that the electro-neutrality of our system is preserved
(except for its surfaces, may be), and no volume charge is formed, so that $%
\nabla _{x}n=0.$ Substituting Eq. (\ref{ex}) to the first of the set of Eqs.
(\ref{j}) one finds that 
\begin{equation}
\widehat{\beta }=\left( 
\begin{array}{cc}
-\frac{\sigma _{xx}}{e}\left( \frac{d\mu }{dT}\right) & c\left( \frac{dM_{z}%
}{dT}\right) \\ 
-c\left( \frac{dM_{z}}{dT}\right) & -\frac{\sigma _{yy}}{e}\left( \frac{d\mu 
}{dT}\right)%
\end{array}%
\right) .  \label{betaxx}
\end{equation}

\bigskip Now we can proceed with obtaining TE and TM coefficients explicitly.

\emph{Seebeck tensor} (differential thermoelectric power) $\widehat{Q}$ is
related to the tensors $\widehat{\beta }$ and $\widehat{\sigma }:$ 

\begin{equation}
\widehat{Q}\left( H\right) =-\widehat{\sigma }^{-1}\left( H\right) \widehat{%
\beta }\left( H\right) =\frac{1}{e}\left( \frac{d\mu }{dT}\right) \widehat{I}%
-c\widehat{\sigma }^{-1}\widehat{e}\left( \frac{d\mathbf{M}}{dT}\right) .
\end{equation}

%
%
%
%
where $\widehat{e}=\left[ 
\begin{array}{cc}
0 & 1 \\ 
-1 & 0%
\end{array}%
\right] $ \ is the Levi-Civitte tensor,$\widehat{I}$ is the identity tensor.

The chemical potential derivative can be found explicitly as:%
\begin{equation}
\frac{d\mu }{dT}=\frac{\partial \mu }{\partial T}+\frac{\partial \mu }{%
\partial n}\frac{dn}{dT}.
\end{equation}

For \textit{the reference case of}\emph{\ }an\emph{\ }isotropic 3D metal
with the parabolic spectrum in zero magnetic field%
\begin{equation}
\frac{\partial \mu }{\partial n}=\frac{\hbar ^{2}}{m}\frac{\pi ^{4/3}}{%
\left( 3n\right) ^{1/3}}
\end{equation}%
and $dn/dT=\nu \partial \mu /\partial T$ with $\nu =mp_{F}/(\pi ^{2}\hbar
^{3})$ as density of states. Using these expressions one finds%
\begin{equation*}
\frac{d\mu }{dT}=2\frac{\partial \mu }{\partial T}.
\end{equation*}%
The temperature derivative of a chemical potential for a degenerate electron
gas is well-known (see, for example, \cite{Abrikosov}):%
\begin{equation}
\mu (T)=\mu (0)-\frac{\pi ^{2}T^{2}}{6}\frac{\nu ^{\prime }\left( \mu
\right) }{\nu \left( \mu \right) }.  \label{mu3}
\end{equation}%
As in the 3D case $\nu \left( \mu \right) \sim \sqrt{\mu \text{, }}$ thus $%
\partial \mu /\partial T=-\pi ^{2}T/(6\mu ),$ and consequently $d\mu /$ $%
dT=-\pi ^{2}T/\left( 3\mu \right) .$This is why we obtain: 
\begin{equation*}
Q(T,0)=\frac{1}{e}\left( \frac{d\mu }{dT}\right) =-\frac{\pi ^{2}T}{3e\mu }.
\end{equation*}%
This expression exactly coincides with the Mott formula for a differential
thermoelectric power \cite{Abrikosov}, that demonstrates the equivalence of
our approach to the classic result obtained the kinetic approach for a 3D
metal.

\begin{figure}[t]
\begin{center}
\includegraphics[ width=1.0\columnwidth ]{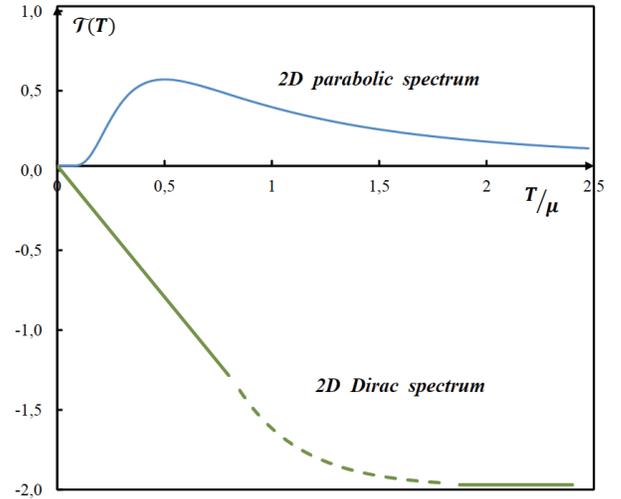}
\end{center}
\caption{Schematic: Thomson coefficient vs temperature in the 2D case. The
Thomson coefficient is positive and non-monotonous for carriers with a
parabolic dispersion, while it is negative and saturates in the case of a
Dirac spectrum.}
\label{thomps}
\end{figure}

\emph{Thomson coefficient}, which describes alternatively heating or cooling
of a current carrying conductor, can be also easily expressed now in terms
of $\mu $ and $M$ temperature derivatives following the Thomson relation 
\begin{equation}
\widehat{\mathcal{T}}=T\frac{d\widehat{Q}}{dT}=\frac{T}{e}\left( \frac{%
d^{2}\mu }{dT^{2}}\right) \widehat{I}-cT\widehat{\sigma }^{-1}\widehat{e}%
\left( \frac{d^{2}\mathbf{M}}{dT^{2}}\right) .
\end{equation}%
%
%
In the absence of magnetic field%
\begin{equation}
\widehat{\mathcal{T}}(T,H=0)=\frac{T}{e}\left( \frac{d^{2}\mu }{dT^{2}}%
\right) \widehat{I}.
\end{equation}%
Using the expressions for the chemical potential summarized in the Table 1
one can see that the Thomson coefficient behaves quite differently for Dirac
and normal carriers. For example, in the degenerate 2D gas of carriers with
a parabolic dispersion the Thomson coefficient is 
\begin{equation}
\mathcal{T}_{(P)}^{\left( 2\right) }=\frac{\left[ \mu _{(P)}^{\left(
2\right) }\left( 0\right) \right] ^{2}}{eT^{2}}\exp \left( -\frac{\mu
_{(P)}^{\left( 2\right) }\left( 0\right) }{T}\right) ,
\end{equation}%
while for 2D Dirac carriers it differs not only by its temperature
dependence but also in its sign: 
\begin{equation}
\mathcal{T}_{(D)}^{\left( 2\right) }=-\frac{\pi ^{2}T}{6e\mu _{D}^{\left(
2\right) }\left( 0\right) }.
\end{equation}%
One can see that in a wide range of temperatures up to $T\lesssim 0.2\mu
_{P}^{\left( 2\right) }\left( 0\right) $\ Thomson coefficient for the normal
carriers is exponentially small. In the same time for Dirac fermions the
Thomson coefficient is of the opposite sign and growth in its absolute value
linearly with temperature. Moreover, at high temperatures the behavior of
the Thomson coefficient is non-monotonous for normal carriers, while for the
Dirac spectrum it is a monotonous saturating function (see Fig. \ref{thomps}%
).

\emph{Nernst-Ettingshausen effect} is the thermal analog of the Hall effect
and it consists in the appearance of an electric field $E_{y}$ perpendicular
to the mutually perpendicular magnetic field $H(\parallel z)$ and
temperature gradient $(\nabla _{x}T)$ (see Fig. \ref{geometry}). It is
characterized by the Nernst coefficient%
\begin{equation}
\nu =\frac{E_{y}}{(-\nabla _{x}T)H},
\end{equation}%
which can be expressed in terms of the resistivity and thermoelectric
tensors \cite{Anselm} : 
\begin{equation}
\nu =-\left( \rho _{xx}\beta _{xy}+\rho _{xy}\beta _{yy}\right) /H.
\label{nurho}
\end{equation}%
Substituting in Eq. (\ref{nurho}) the expressions for the thermoelectric
tensor components obtained above, we arrive at: 
\begin{equation}
\nu =\frac{\sigma _{xx}}{e^{2}nc}\left( \frac{d\mu }{dT}\right) +\frac{c\rho
_{yy}}{H}\left( \frac{dM_{z}}{dT}\right) .  \label{nu}
\end{equation}%
The first term here is governed by the temperature dependence of the
chemical potential, while the second is related to magnetization currents.
In our reference case of a 3D metal the second term is negligible by a
parameter $\left( k_{F}l\right) ^{-1}\ll 1$, with $k_{F}$ being the Fermi
wave-vector, $l$ being the mean free path. Using Eq. (\ref{mu3}) one easily
reproduces the well-known Sondheimer formula \cite{1948Sondheimer}

\begin{equation}
\nu =\frac{\sigma _{xx}}{e^{2}nc}\left( \frac{d\mu }{dT}\right) =-\frac{%
\sigma _{xx}}{e^{2}nc}\frac{\pi ^{2}T}{3\mu }=-\frac{\pi ^{2}T}{3\mu }\frac{%
\tau }{mc}.
\end{equation}%
In fluctuating superconductors the role of the second term in Eq. (\ref{nu})
becomes crucial: it \textquotedblleft saves\textquotedblright\ the third law
of thermodynamics in the vicinity of the second critical field $H_{c2}(0)$ 
\cite{SSVG09}. The oscillations of Nernst-Ettingshausen coefficient in
graphene obtained within the present approach \cite{VarlamovPRL} demonstrate
a remarkable agreement with experimental data \cite{Zuev}.

In the rest of this Letter we summarize the useful expressions for other
important TE and TM coefficients following the approach formulated here.

\emph{Peltier tensor} which describes the heat generation by electric
current, is given by Kelvin relation (see e.g. \cite{Abrikosov}) and can
also be expressed in terms of conductivity and thermoelectric tensors

\begin{equation}
\widehat{\Pi }\left( T,H\right) =T\widehat{Q}\left( T,H\right) =-T\widehat{%
\sigma }^{-1}\left( H\right) \widehat{\beta }\left( H\right) .
\end{equation}

At zero magnetic field, it has only diagonal components:

\begin{equation}
\Pi \left( T,0\right) =\frac{T}{e}\left( \frac{d\mu }{dT}\right) .
\end{equation}

Interestingly, while the whole system is out of thermal equilibrium in the
presence of electric current, the Peltier coefficient can still be linked to
the thermal derivative of the chemical potential!

\emph{The thermal conductivity tensor} $\widehat{\kappa }$ describes the
ability of a material subject to a temperature gradient to conduct heat. It
can be expressed through the elements of TE tensor and electric conductivity

\begin{equation}
\widehat{\kappa }=-\frac{\pi ^{2}T}{3e^{2}}\widehat{\sigma }\left( \widehat{I%
}-\frac{3e^{2}}{\pi ^{2}}\left[ \widehat{\beta }\widehat{\sigma }^{-1}\right]
^{2}\right) .
\end{equation}

\emph{Righi-Leduc effect} describes the heat flow resulting from a
perpendicular temperature gradient in the absence of electric current.
Righi-Leduc coefficient can be expressed through the diagonal elements of
thermal conductivity and conductivity tensors as 
\begin{equation}
L=\frac{\sigma _{xx}}{enc}\kappa _{yy}.
\end{equation}

Finally, we would like to discuss limitations of our approach. We have
largely used the electro-neutrality condition $\nabla _{x}n=0$, which may
fail in certain semiconductor systems where the volume charge effects are
important. For the same reason, this approach fails to account for electric
currents induced by a phonon drag. 

In conclusion, the crucial function which governs all TE and TM constants
listed above is the temperature derivative of the chemical potential $\frac{%
d\mu }{dT}$. This simple observation opens way to the prediction of TE and
TM effects in new structures and materials. In particular, it shows that TE
and TM\ properties may be strongly different in systems with Dirac fermions
and normal carriers. 

The authors acknowledge helpful discussions with I. Chikina and S.G.
Sharapov. This work has been supported by the EU\ IRSES program "SIMTECH".

\end{document}